\documentstyle[aps,prl,multicol]{revtex}      

\begin{document}
\topmargin=-0.5in
\input{epsf}


\title{The relationship between fragility, configurational entropy and
the potential energy landscape of glass forming liquids}

\author{Srikanth Sastry\\
Jawaharlal Nehru Centre for Advanced Scientific Research,Jakkur Campus, Bangalore 560064, INDIA}
\maketitle

\noindent
{\bf Glass is a microscopically disordered, solid form of matter that
results when a fluid is cooled or compressed in such a fashion that it
does not crystallise. Almost all types of materials are capable of
glass formation -- polymers, metal alloys, and molten salts, to name a
few. Given such diversity, organising principles which systematise
data concerning glass formation are invaluable. One such principle is
the classification of glass formers according to their
fragility\cite{fragility}. Fragility measures the rapidity with which
a liquid's properties such as viscosity change as the glassy state is
approached. Although the relationship between features of the energy
landscape of a glass former, its configurational entropy and fragility
have been analysed previously (e. g.,\cite{speedyfr}), an
understanding of the origins of fragility in these features is far
from being well established. Results for a model liquid, whose
fragility depends on its bulk density, are presented in this letter.
Analysis of the relationship between fragility and quantitative
measures of the energy landscape (the complicated dependence of energy
on configuration) reveal that the fragility depends on
changes in the vibrational properties of individual energy basins, in
addition to the total number of such basins present, and their spread
in energy. A thermodynamic expression for fragility is derived, which
is in quantitative agreement with {\it kinetic} fragilities obtained
from the liquid's diffusivity.}

	Glass forming liquids grow increasingly viscous upon cooling,
till the viscosity becomes so large that they fail to flow on
experimental time scales. While remaining microscopically disordered
like a liquid, they manifest mechanical properties of a solid; {\it
i. e.}, they transform to a glass. By convention, the {\it glass
transition} temperature $T_g$ is that where the viscosity reaches a
value of $10^{12}$ Pa sec. The approach to this large viscosity,
however, differs from one liquid to another. When displayed in an
Arrhenius plot [log(viscosity) {\it vs.} inverse temperature $1/T$],
some liquids ({\it e. g.}, silica) show a steady, linear increase,
while others display a much steeper dependence on
$1/T$\cite{fragility,laughlin}, as illustrated in Fig. 1(a)
(inset). The former are {\it strong} liquids, and the latter, {\it
fragile}. This range of behaviours is implicit in the
Vogel-Fulcher-Tammann-Hesse (VFT) form, observed to describe the $T$
dependence of viscosity (as well as diffusivity and relaxation times)
in many glass formers. The VFT relation may be written as
\begin{equation} \label{eq:vft}
\eta = \eta_0 \exp\left[{1\over K_{\small VFT}(T/T_0 - 1)}\right], 
\end{equation}
where $T_0$ is the temperature of apparent divergence of viscosity,
and $K_{\small VFT}$ is a material specific parameter quantifying the {\it
kinetic} fragility; more fragile liquids have larger $K_{\small VFT}$ values.
This behaviour of viscosity correlates with the jump in heat
capacity at the glass transition -- more fragile the liquid,  sharper
and bigger the jump. Rationalisation of this correlation comes from
the Adam-Gibbs relationship\cite{adam-gibbs} which predicts a
dependence of viscosity on the configurational entropy $S_c$
(described later) of the liquid:
\begin{equation} \label{eq:ag}
\eta = \eta_0 \exp\left[{A\over T S_c}\right], 
\end{equation}
which results in the VFT relation if $S_c$ has the form $T S_c =
K_{\small AG}(T/T_K - 1)$\cite{tkto}, where $T_K$ is the ideal glass
transition or Kauzmann temperature\cite{adam-gibbs} below which $S_c =
0$. The jump in heat capacity is proportional to $K_{\small AG}$, and
is larger for more fragile glass formers.

Fragility is analysed here for a binary mixture of particles
interacting {\it via} the Lennard-Jones potential, widely studied as a
model glass former\cite{kob,sastrynature,fs,parisi,sastryprl}. The $T$
dependence of diffusivity $D$ (to characterise dynamics) and
configurational entropy are obtained for a range of bulk densities $\rho$.
The diffusivities $D(\rho, T)$, are shown in Fig. 1(a) in a `fragility
plot': $-\log[D(\rho, T)]$ {\it vs.}  $T_r/T$, where $T_r$ is a {\it
reference} temperature (akin to $T_g$ in the usual fragility plot)
at which $D$ at each density reaches a fixed, small value.  By
comparison with the inset, the liquid is seen to
become more fragile as its density increases. Fig. 1(b) shows the
kinetic fragility index, $K_{\small VFT}$, obtained from VFT
(Eq.(\ref{eq:vft})) fits to diffusivity data at each density.  The
density dependence of $K_{\small VFT}$ quantifies the increase of fragility
with density, apparent from data in Fig. 1(a).

A thermodynamic explanation of fragility, as discussed above (also
\cite{ito,wolynes}), lies in explaining the rapidity of change with
$T$ of the configurational entropy, $S_c$. In the inherent structure
(IS) formalism\cite{FHSINH,FHS95} employed here, the configuration
space of a liquid is divided into basins of local potential energy
minima (`inherent structures'). $S_c$ at a given $T$ derives from the
multiplicity of local potential energy minima sampled by the liquid at
that $T$, and equals the difference of the total entropy and the
vibrational entropy of typical basins sampled, $S_{vib}$:

\begin{equation} \label{eq:scdef}
S_c(\rho,T) = S_{total}(\rho,T) - S_{vib}(\rho,T).
\end{equation}

For all the results presented here, the basin entropy $S_{vib}$ is
calculated by approximating each basin as a harmonic well (valid at
sufficiently low $T$)\cite{fs,parisi,sastryprl,heuer}. In this
approximation, the entropy of a given basin, arising from vibrational
motion within the basin (the suffix {\it vib} emphasizes
this point) is given by
\begin{equation} \label{eq:sbdef}
S_{vib} = k_B \sum_{i = 1}^{3N} 1 - \log({h \nu_i \over  k_B T}),
\end{equation}
where $\nu_i$ are the vibrational frequencies of the given basin,
$k_B$ is Boltzmann's constant and $h$ is Plank's constant. The basin
free energy is $F_{vib} = U_{vib} - T S_{vib}$, where $U_{vib} = 3 N
k_B T$ is the internal energy. Apart from the explicit temperature
dependence of $S_{vib}$ above, an implicit dependence also exists
because vibrational frequencies change from one basin to another,
and, different basins are sampled at different
temperatures. From the expression above, it is clear that the entropy
difference between any two basins arises solely from the difference in
their vibrational frequencies, and is independent of temperature.

The configurational entropy is evaluated in this letter using two
methods, referred to as the thermodynamic integration (TI) method, and
the potential energy landscape (PEL) method, respectively. In both
cases, inherent structures are obtained by performing local potential
energy minimisation for a subset of configurations sampled by a liquid
at each $\rho$ and $T$. Vibrational frequencies are calculated for
each inherent structure, and the basin entropy $S_{vib}$ is obtained
from Eq. (\ref{eq:sbdef}). In the TI method (described in detail
elsewhere\cite{fs,parisi,sastryprl}; see also caption of Fig. 2), the
total entropy $S_{total}$ is obtained using pressure and internal
energy values from simulations. Eq. (\ref{eq:scdef}) is used to obtain
$S_c$. In contrast, the PEL method is based on constructing the
distribution of inherent structures in order to calculate $S_{total}$ and
$S_c$, as described later.

Fig. 2(a) shows the Adam-Gibbs plot, where $\log D$ is plotted
against $(TS_c)^{-1}$. The Adam-Gibbs relation (Eq. (\ref{eq:ag})) is
seen to apply well at each density, as also observed previously for
the hard sphere liquid\cite{speedyag} and water\cite{scala}.
Fig. 2(b) shows (as lines) the $T$ dependence of $TS_c$ (from TI).
All curves are nearly linear, but with slight increases of slope
at lower $T$. The slopes of $TS_c$ (in the $T$ range where diffusivities
are measured), $K^{\small{\small TI}}_{\small AG}$, are used as the quantitative
index of thermodynamic fragility, and are plotted in
Fig. 1(b). $K^{\small{\small TI}}_{\small AG}$ are nearly the same as the kinetic
fragility index $K_{\small VFT}$, thereby validating the idea that the rate
of increase of $T~S_c$ does indeed determine the fragility.

However, this analysis provides no insight into how the fragility
relates to features of the energy landscape. To this end, the
distribution of energy minima, and properties of individual basins,
are now examined, at fixed density $\rho$. The probability
distribution $P(\Phi,T)$ that an inherent structure of energy $\Phi$
is sampled at temperature $T$ depends on (a) the energy $\Phi$, (b)
the structure of the basin, contained in the basin free energy
$F_{vib}(\Phi,T)$ and (c) the number density of inherent structures
with energy $\Phi$,
$\Omega(\Phi)$\cite{FHSINH,FHS95,fs,heuer,sastrytrieste}:
\begin{equation} \label{eq:pofe}
P(\Phi,T) = \Omega(\Phi) ~exp\left[-\beta (\Phi + F_{vib}(\Phi,T))\right]/Q_N(\rho,T),
\end{equation} 
where $Q_N$, the partition function, normalises $P$ and is given by
$Q_N = \int d\Phi~ P(\Phi,T)$ over all possible $\Phi$.  The
configurational entropy density is defined by ${\cal S}_c(\Phi)
\equiv k_B \ln \Omega(\Phi)$ and is related to the $T$ dependent
configurational entropy by $S_c(T) = \int d\Phi~ {\cal S}_c(\Phi)
P(\Phi,T)$. The quantities $P(\Phi,T)$, $F_{vib}$ and $Q_N$ are
obtained from simulations\cite{fs,heuer,sastryprl,sastrytrieste}, and
are used to invert the relation in Eq. (\ref{eq:pofe}) to obtain the
configurational entropy density ${\cal S}_c(\Phi)$. The same ${\cal
S}_c(\Phi)$ estimates must result (though not for the same range of
$\Phi$) for different $T$, when the harmonic assumption is valid,
which is indeed seen to be the case from data shown in
Fig. 3(a). With increasing density, the maximum ${\cal S}_c(\Phi)$
value reached decreases, and the distribution becomes broader. It has been
argued\cite{speedydist,heuertrieste} that the distribution $\Omega(\Phi)$
should be Gaussian (equivalently, ${\cal S}_c(\Phi)$ an inverted
parabola). Data in Fig. 3(a) are fully consistent with this
expectation, and are fitted to the form,
\begin{equation}\label{eq:scdos} 
{{\cal S}_c(\Phi) \over N k_B} = \alpha - {(\Phi - \Phi_o)^2\over \sigma^2}
\end{equation} 
where $\alpha$ is the height of the parabola and determines the total
number of configurational states, {\it i. e.} energy minima (the total
number is proportional to $\exp(\alpha N)$ ), $\Phi_0$ and $\sigma^2$
respectively define the mean and the variance of the distribution. The
fits are displayed in Fig. 3(a) and the fit parameters listed in the
figure caption. Speedy\cite{speedyfr} has considered a model of the
thermodynamics of a glass former with a Gaussian distribution of
configurational states, and the basin specific heat that does not
depend on the {\it internal parameter} (here the IS energy). He
predicts that the fragility of a liquid should be proportional to the
logarithm of the number of states, {\it i. e.} on $\alpha$. From
Fig. 3(a) it is clear that $\alpha$ values, and hence the number of
states, decrease with increasing density $\rho$. But the fragilities
(Fig. 1(b)) show the opposite trend. However, fragility has an additional 
dependence (implicit in ref. \cite{speedyfr}) on the variance of the distribution $\sigma$ (see equation (6)).

Further, the variation
of vibration frequencies, and hence the basin entropy, from basin to
basin, also contributes to the fragility of the liquid. Fig. 3(b)
displays the $T$ independent (see above) basin entropy difference at a
given energy $\Phi$ with respect to the value at $\Phi_0$, $\Delta
S_{vib}(\Phi) \equiv S_{vib}(\Phi,T) - S_{vib}(\Phi_0,T)$.  At all
densities, $\Delta S_{vib}(\Phi)$ displays a clear $\Phi$ dependence,
which is inconsistent with assumptions in \cite{speedyfr}. Although
such basin dependence of vibrational frequencies and hence the basin
entropy have been discussed
previously\cite{goldstein,kjrao,angelltrieste,jenny,johari1,johari2,yamamuro,ngai,fsaging,starrwater},
with reference to both experimental data and model calculations, its
implications on a liquid's fragility have not clearly been
established.  Fig. 3(b) shows that basins at higher energy (which are
occupied at higher temperature) have lower basin entropies, or, higher
frequencies (see Eq. (\ref{eq:sbdef})). This trend, contrary to what
one expects for systems studied experimentally, arises because the
constant density model system here cannot expand when $T$ is raised.

The dependence of $\Delta S_{vib}(\Phi)$ on $\Phi$ is to a very good
extent linear. Hence, for each density, $\Delta S_{vib}$ is fit to the
form $\Delta S_{vib} (\Phi) = \delta S~ (\Phi - \Phi_0)$ with fit
parameters listed in the caption of Fig. 3. The basin free energy,
which can be written as $F_{vib}(\Phi,T) = F_{vib}(\Phi_0,T) - T
\delta S (\Phi - \Phi_0)$, is also linear in $\Phi$. With a Gaussian
form for $\Omega$, the partition function $Q_N = \int d\Phi P(\Phi,T)$
($P$ as in Eq. (\ref{eq:pofe})) is easily evaluated, and
results in the following predictions for the $T$ dependence of the
average IS energy, and the configurational entropy:
\begin{equation} \label{eq:phi}
<\Phi> (T) = \Phi_0^{eff} - {\sigma^2 \over 2 N k_B T},
\end{equation} 
where $\Phi_0^{eff} = \Phi_0 + {\sigma^2 \delta S \over 2 N k_B}$, and 
\begin{equation} \label{eq:kagPEL}
T S_c(T) = K^{\tiny{PEL}}_{\small AG}(T)~~ (T/T_K -1); ~~~K^{\small{\small PEL}}_{\small AG}(T) = \left({\sigma \sqrt{\alpha} \over 2} + {\sigma^2 \delta S\over 4 N k_B}\right) \left( 1 + {T_K\over T}\right) - {\sigma^2 \delta S\over 2 N k_B},
\end{equation}
where $T_K = \sigma(2 N k_B\sqrt\alpha + \sigma \delta S)^{-1}$ is the ideal glass transition temperature where $S_c(T_K) = 0$.

The form above for $S_c$ results in the VFT relation, if
$K^{\small{\small PEL}}_{\small AG}(T)$ is constant. This is not quite the case, but
the $1/T$ term becomes rapidly irrelevant for $T > T_K$ and
$K^{\small{\small PEL}}_{\small AG}(T)$ approaches a constant. Further, $T S_c$
values plotted (as points) in Fig. 2(b) are in extremely good
agreement with the ones from thermodynamic integration, including
deviations from linearity for small $T/T_K$. A check of the overall
consistency of the model described here is made in Fig. 3(c), which
shows a {\it scaled plot} of $<\Phi>(T) - \Phi_0^{eff}$ {\it vs.}
${\sigma^2 \over 2 N k_B T}$.  From Eq. (\ref{eq:phi}), one expects $\Phi(T)$
data for all densities to collapse onto a straight line of (negative)
unit slope, with expected deviations at high $T$ due to basin
anharmonicity\cite{heuer}. Data in Fig. 3(c) clearly meet this expectation,
although with noticeable deviations for $\rho = 1.35 \rho_0$, as also in
Fig. 2(b).

The expression for $K^{\small{\small PEL}}_{\small AG}$ in
Eq. (\ref{eq:kagPEL}), the thermodynamic fragility index obtained from
energy landscape parameters, shows that fragility depends not only on
the multiplicity of states ($\alpha$) but also on their spread
($\sigma$), and how much the basin entropy changes from the lowest to
the highest energy basins sampled by the liquid. The extent of this
change is quantified by parameters $\sigma$ and $\delta S$ (listed in
the caption of Fig. 3). Fig. 1(b) shows the average slopes of $TS_c$
(from Eq. (\ref{eq:kagPEL}) above, in the $T$ range where
diffusivities are measured), $K^{\small PEL}_{\small AG}$, which are
nearly the same as $K^{\small TI}_{\small AG}$, and in very good
agreement with the kinetic fragility index $K_{\small{VFT}}$. This
agreement establishes the {\it quantitative} relationship between
fragility and the energy landscape of the liquid since
$K^{\small{\small PEL}}_{\small AG}$ is obtained solely in terms of a
quantitative description of the liquid's energy landscape.

\noindent{\bf Acknowledgements:} I thank C. A. Angell, G. P. Johari,
K. J. Rao, F. Sciortino, R. Seshadri, R. J. Speedy and U. V. Waghmare
for very useful discussions and/or comments on the manuscript.

\medskip

\leftline{Correspondence should be addressed to the authour (email: sastry@jncasr.ac.in).} 
\eject

\begin{figure}[h]
\end{figure}

{\bf Figure 1:} Fragility Plot of diffusivities, kinetic and
thermodynamic fragility indices. Results shown are obtained from
molecular dynamics simulations of $204$ $A$ type and $52$ $B$ type
particles interacting {\it via} the Lennard-Jones (LJ)
potential. Argon units are used for $A$ type particles: LJ parameter
$\epsilon_{AA} = k_B \times 119.8 K$, $\sigma_{AA} = 0.3405 \times
10^{-9} m$, $m_{A} = 6.6337 \times 10^{-26} kg$, and
$\epsilon_{AB}/\epsilon_{AA} = 1.5$, $\epsilon_{BB}/\epsilon_{AA} =
0.5$, $\sigma_{AB}/\sigma_{AA} = 0.8$, and $\sigma_{BB}/\sigma_{AA} =
0.88$, $m_{B}/m{A} = 1.$ Densities are reported in units of $\rho_0 =
2.53 \times 10^{28} m^{-3}$ or $1.678 g cm^{-3}$. The liquid-gas critical
point is located roughly at $T_c = 130 K$, $\rho_c = 0.416 \rho_0$.
(a) Diffusivities $D$ from molecular dynamics simulations (details of
the simulations are as in $\cite{sastryprl}$) are displayed in a
`fragility plot' for $5$ densities. The reference temperature $T_r$ is
chosen such that $D(T_r) = 2.44 \times 10^{-8} cm^2/s$, which is
slightly below the lowest $D$ values measured. The VFT extrapolation
is used to locate $T_r$. Fit values $D_0$ (preexponent) are in the
range of $0.87$ to $2.0 \times 10^{-4} cm^2/s$, close to the
experimentally observed values. (b) Fragility index $K$ obtained (i) from VFT
fits to diffusivities $D$, and from the configurational entropy
obtained (ii) from thermodynamic integration (`TI'), and (iii) from
analysis of the potential energy landscape (`PEL'). Dimensionless
thermodynamic fragility indices are obtained by dividing $K^{\small
TI}_{\small AG}$, $K^{\small PEL}_{\small AG}$ by
$N\epsilon_{AA}$. Comparison with data in $\cite{ito}$ shows
that at the low density end, the fragility of the model liquid
compares with that of considerably fragile liquids such as toluene,
orthoterphenyl and salol, while at the high density end, the fragility
is extremely high. Temperatures of vanishing diffusivity $T_0$ from
VFT fits are $~18.76 K$, $~24.98 K$, $~35.97 K$, $~47.86 K$ and $~75.75 K$
for $\rho/\rho_0 = ~1.1$, $~1.15$, $~1.2$,$~1.25$ and $~1.35$ respectively.

\medskip

{\bf Figure 2:} Adam-Gibbs plot, configurational entropy {\it vs.} temperature.
(a) The `Adam-Gibbs' plot displaying $\log(D)$ {\it vs.}
$(T~S_c)^{-1}$. $T~S_c$ is expressed in units of $\epsilon_{AA}$.  The
Adam-Gibbs expectation of a linear dependence is well satisfied at
each density. (b) Configurational entropy obtained from thermodynamic
integration (`TI') (lines, not labeled), and from analysis of the
potential energy landscape (points, labeled as `PEL'), plotted against
temperature $T$ scaled to the Kauzmann temperature $T_K$ obtained
from thermodynamic integration. In the TI method, the total entropy
$S_{total}$ of the liquid is obtained from its total free energy $F_{total}$, from $F_{total}
= U_{total} - T S_{total}$, where $U_{total}$ is the internal energy. Thermodynamic identity
$P = \rho^2 \left({\partial F_{total}/N\over \partial \rho}\right)_T$ where
$P$ is the pressure, is used to integrate $P$ from simulations to
obtain $F_{total}$ with respect to the known ideal gas free energy. The
identity $U_{total} = \left({\partial F_{total}/T \over \partial 1/T}\right)_\rho$ is
used to integrate at fixed density $U_{total}$ from simulations to obtain $F_{total}$
at any desired $T$.  Extrapolations to low $T$ beyond the range of
simulations utilize the form for the potential energies $E$, $E = E_0
+ E_1 T^{E_2}$, where $E_2 \sim 3/5$ (see \cite{fs,parisi,sastryprl}). Basin entropies $S_{vib}$ are obtained by
averaging over typical inherent structures at each $T$, whose
vibrational frequencies are obtained numerically by diagonalising the
Hessian (matrix of second derivatives of the potential energy). The
plot covers at each density the range of $T$ where simulations have
been performed. Scaling temperatures $T^{\small TI}_K$, the
Kauzmann or ideal glass transition temperature where $S_c = 0$, are:
$20.44 K, ~26.83 K, ~34.35 K, ~42.49 K$, and $~62.47 K$, for $\rho/\rho_0 =
~1.1, ~1.15, ~1.2,~1.25$ and $~1.35$ respectively. $T^{\small
PEL}_K$ obtained from the PEL method are, in the same order, $20.24
K, ~26.18 K, ~34.74 K, ~44.64 K$, and $~62.35 K$.

{\bf Figure 3:} Distribution of energy minima, basin entropy, and
temperature dependence of inherent structure energies. (a) The
configurational entropy density ${\cal S}_c(\Phi)$ plotted against
inherent structure energy $\Phi$ minus the `Kauzmann' energy
$\Phi_{min}$ below which ${\cal S}_c(\Phi) = 0$. $\Phi_{min}$ is the
lower root of ${{\cal S}_c(\Phi) \over N k_B} = \alpha - {(\Phi -
\Phi_o)^2\over \sigma^2} = 0.$ For each density, data from $9$
temperatures (on average) are shown (smaller symbols) which collapse
onto each other, validating the harmonic approximation for calculating
basin entropies (see text). Also shown are the averages of these
curves (larger symbols), for whose calculation a threshold of $20$
samples (out of a total of $1000$) is imposed in each bin. (b) The
$T$-independent component of the basin entropy $\Delta S_{vib}(\Phi)$
{\it vs.} $\Phi - \Phi_o$, where $\Phi_o$ is the value where ${\cal
S}_c(\Phi)$ is maximum. $\Delta S_{vib}(\Phi)$ curves are
approximately linear.  The $T$ independence of $\Delta S_{vib}(\Phi)$
is an artefact of the classical mechanical treatment employed here,
but a quantum mechanical calculation of the basin entropy does not
alter the picture fundamentally (unpublished results). (c) Scaled plot
of the average IS energies sampled in simulations, which are expected
to collapse on to the straight line shown at low temperatures (right
side of the graph), if the assumption of a Gaussian configurational
entropy density ${\cal S}_c(\Phi)$ is valid (see text). The observed
data collapse validates the Gaussian form used for ${\cal
S}_c(\Phi)$. Deviations at high temperatures are due to the breakdown
of harmonicity of basins. Fit parameters $\alpha$,
$\phi_0/N\epsilon_{AA}$, $\sigma/N\epsilon_{AA}$ and $\delta S
\epsilon_{AA}/k_B$ are, ($\rho = 1.1 \rho_0$): $1.010,~ -6.586,
~0.303, ~-0.721,$ ($\rho = 1.15 \rho_0$): $~0.957, ~-6.684, ~0.374,
~-0.652,$ ($\rho = 1.2 \rho_0$): $~0.933, ~-6.683, ~0.489, ~-0.500,$
($\rho = 1.25 \rho_0$): $~0.927, ~-6.564, ~0.642, ~-0.314,$ and ($\rho
= 1.35 \rho_0$): $~0.849, ~-6.088, ~0.873, ~-0.188.$

\eject 
\centerline{\large\bf SASTRY FIGURE 1}
\vfill
\begin{figure}[h]
\hbox to\hsize{\epsfxsize=1.0\hsize\hfil\epsfbox{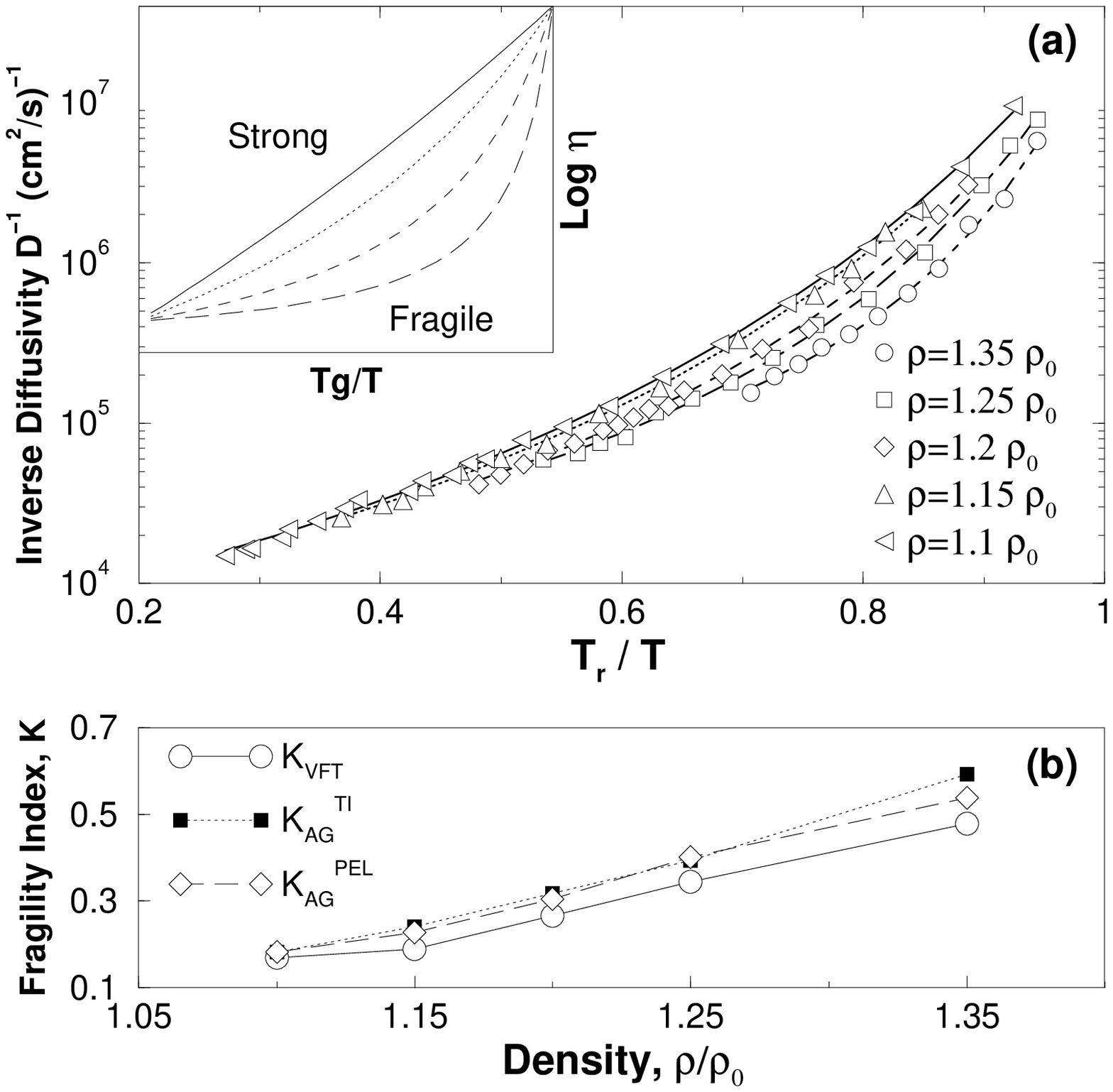}\hfil}
\end{figure}
\vfill
\eject

\centerline{\large\bf SASTRY FIGURE 2}
\vfill
\begin{figure}[h]
\hbox to\hsize{\epsfxsize=1.0\hsize\hfil\epsfbox{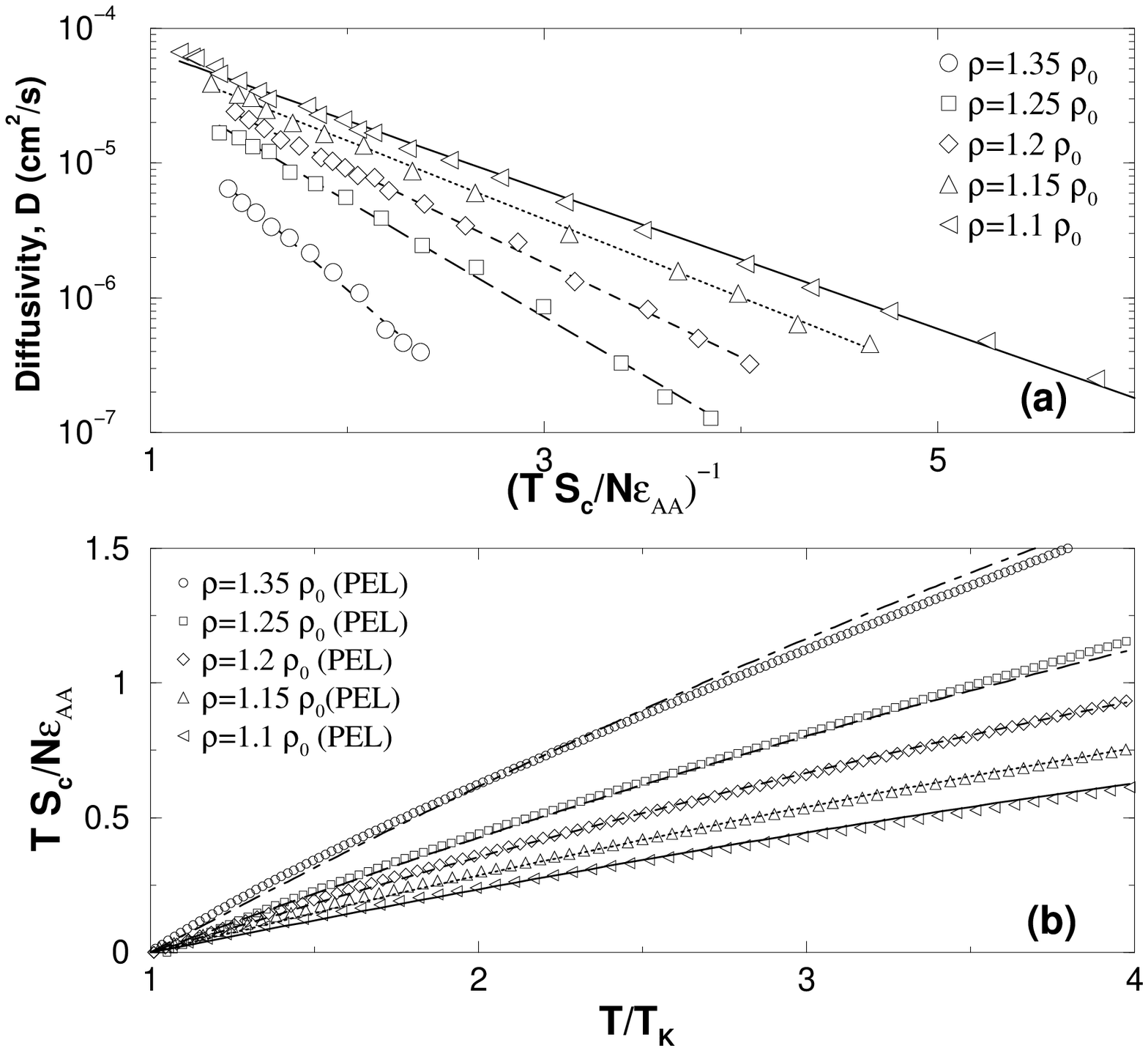}\hfil}
\end{figure}
\vfill
\eject

\centerline{\large\bf SASTRY FIGURE 3}
\vfill
\begin{figure}[h]
\hbox to\hsize{\epsfxsize=1.0\hsize\hfil\epsfbox{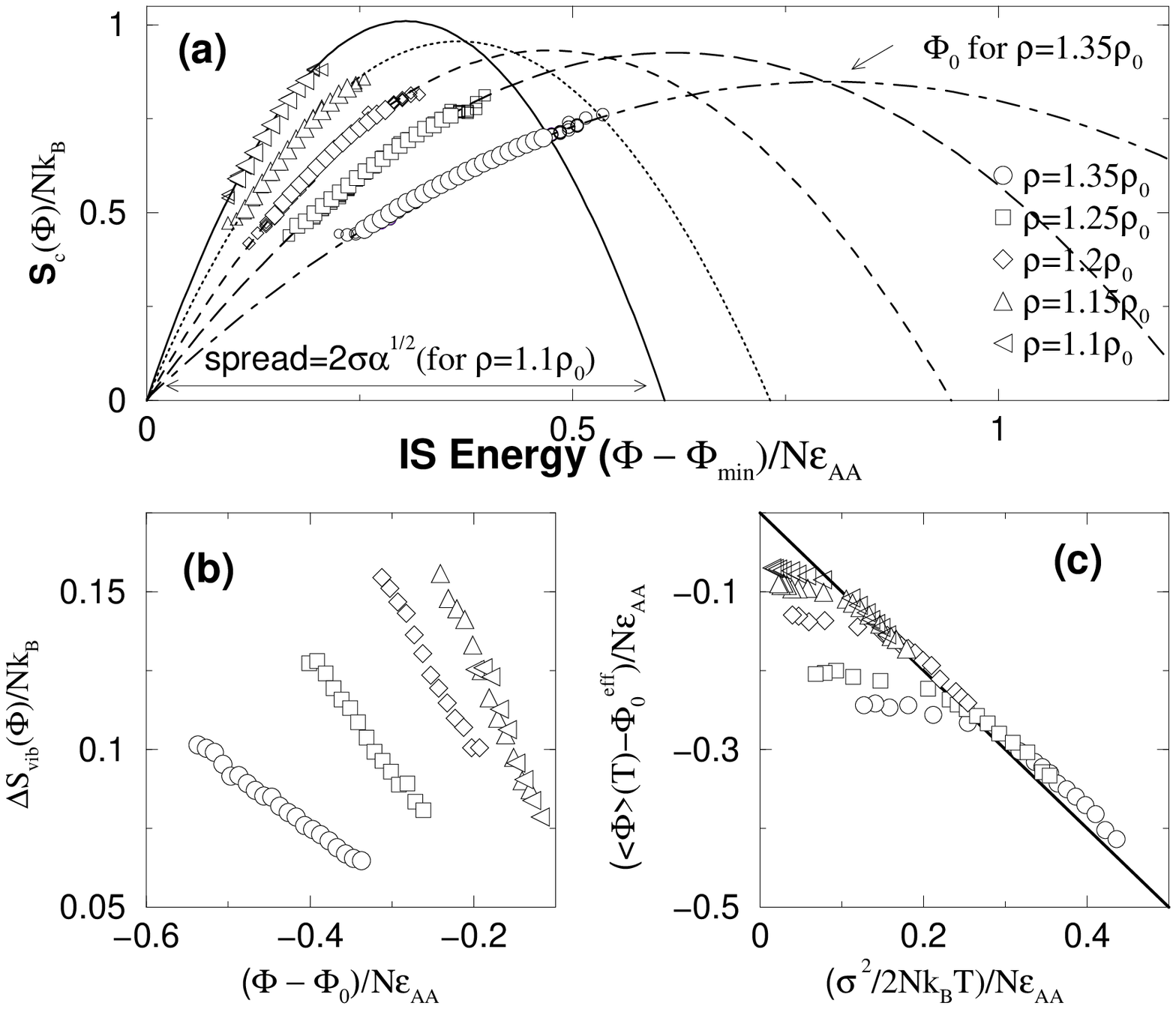}\hfil}
\end{figure}
\vfill


\medskip 

\begin{thebibliography}{999}

\bibitem{fragility} Angell, C. A. Relaxation in Liquids,
Polymers and Plastic Crystals -- Strong/Fragile Patterns and
Problems. {\it J. Non-Cryst. Solids} {\bf 131-133}, 13--31 (1991).

\bibitem{speedyfr} Speedy, R. J. Relations between a Liquid and Its
Glasses. {\it J. Phys. Chem. B} {\bf 103}, 4060--4065 (1999).

\bibitem{laughlin} Laughlin, W. T. and Uhlmann, D. R. Viscous Flow in
Simple Organic Liquids. {\it J. Phys. Chem.} {\bf 76}, 2317--2325 (1972).

\bibitem{adam-gibbs} Adam, G.  and Gibbs, J. H On the Temperature Dependence
of Cooperative Relaxation Properties in Glass-Forming Liquids. {\it
J. Chem. Phys.} {\bf 43}, 139--146 (1965).

\bibitem{kob} Kob, W. and Andersen, H. C. Testing mode-coupling theory
for a supercooled binary Lennard-Jones mixture: The van Hove
correlation function. {\it Phys. Rev. E} {\bf 51}, 4626--4641 (1995).

\bibitem{sastrynature} Sastry, S., Debenedetti, P. G. and Stillinger,
F. H. Signatures of Distinct Dynamical Regimes in the Energy Landscape
of a Glass Forming Liquid. {\it Nature} {\bf 393}, 554--557 (1998).

\bibitem{fs} Sciortino, F., Kob, W. and Tartaglia, P. Inherent
Structure Entropy of Supercooled Liquids.  {\it Phys. Rev. Lett.} {\bf
83}, 3214--3217 (1999).

\bibitem{parisi} Coluzzi, B., Parisi G. and Verrocchio,
P. Lennard-Jones binary mixture: a thermodynamical approach to glass
transition.  {\it J. Chem. Phys.} {\bf 112} 2933--2944 (2000).

\bibitem{sastryprl} Sastry, S. Liquid Limits: The Glass Transition and
Liquid-Gas Spinodal Boundaries of Metastable Liquids. {\it
Phys. Rev. Lett.} {\bf 85} 590--5593 (2000).

\bibitem{ito} Ito, K., Moynihan, C. T. and Angell, C. A. Thermodynamic
determination of fragility in liquids and a fragile-to-strong liquid
transition in water. {\it Nature} {\bf 398}, 492--495 (1999).

\bibitem{wolynes} Xia, X. and Wolynes, P. G. Fragilities of Liquids 
Predicted from the Random First Order Transition Theory of Glasses. {\it 
Proc. Nat. Acad. Sci.} {\bf 97} 2990-2994 (2000). 

\bibitem{FHSINH} Stillinger, F. H. and  Weber, T. A. Packing Structures
and Transitions in Liquids and Solids. {\it Science} {\bf 225}, 983-989
(1984).

\bibitem{FHS95} Stillinger, F. H. A Topographic View of Supercooled
Liquids and Glass Formation. {\it Science} {\bf 267}, 1935--1939 (1995).

\bibitem{heuer} Buechner, S. and Heuer, A. The potential energy landscape 
of a model glass former: thermodynamics, anharmonicities, and finite size 
effects. {\it Phys. Rev. E} {\bf 60}, 6507--6518 (1999).

\bibitem{speedyag} Speedy, R. J. The hard sphere glass
transition. {\it Mol. Phys.} {\bf 95} 169-178 (1998).

\bibitem{scala} Scala, A., Starr, F. W., La Nave, E., Sciortino, F. and
Stanley, H. E. Configurational Entropy and Diffusivity of Supercooled Water.
{\it Nature} {\bf 406} 166--169 (2000).

\bibitem{sastrytrieste} Sastry, S. Evaluation of configurational
entropy of a model liquid from computer simulations.  Proceedings of
{\it Unifying Concepts in Glass Physics, Trieste, 1999}, {\it
J. Phys. Cond. Mat.} {\bf 12}, 6515--6524 (2000).

\bibitem{speedydist} Speedy, R. J. and Debenedetti, P. G. The
distribution of tetravalent network glasses {\it Mol. Phys.} {\bf 88}
1293--1316 (1996).

\bibitem{heuertrieste} Heuer, A. and Buechner, S. Why is the density
of inherent structures of a Lennard-Jones type system gaussian?
Proceedings of {\it Unifying Concepts in Glass Physics, Trieste,
1999}, {\it J. Phys. Cond. Mat.} {\bf 12}, 6535--6543 (2000).

\bibitem{tkto} Angell, C. A. Entropy and Fragility in Supercooled
Liquids.  {\it J. Res. NIST} {\bf 102}, 171--185 (1997).

\bibitem{goldstein} Goldstein, M. Viscous liquids and the glass
transition. V. Sources of the excess specific heat of the liquid. {\it
J. Chem. Phys.} {bf 64}, 4767--4774 (1976).

\bibitem{kjrao} Angell, C. A. and Rao, K. J. Configurational Excitations 
in Condensed Matter and the Bond Lattice Model for the Liquid-Glass Transition. {\it J. Chem. Phys.} {bf 57}, 470--481 (1972).

\bibitem{angelltrieste} Angell, C. A. Ten questions on glassformers,
and a real space ``excitations'' model with some answers on fragility
and phase transitions. Proceedings of {\it Unifying Concepts in Glass
Physics, Trieste, 1999}, {\it J. Phys. Cond. Mat.}{\bf 12}, 6463--6476 (2000).

\bibitem{jenny} Green, J. L., Ito, K., Xu, K. and Angell, C. A. Fragility 
in Liquids and Polymers: New, Simple Quantifications and Interpretations. 
{\it J. Phys. Chem. B} {\bf 103}, 3991--3996 (1999).  

\bibitem{johari1} Johari, G. P. A resolution for the enigma of a
liquid's configurational entropy-molecular kinetics relation. {\it
J. Chem. Phys.}  {\bf 112} 8958--8969 (2000)

\bibitem{johari2} Johari, G. P. Contributions to the entropy of a
glass and liquid, and the dielectric relaxation time. {\it
J. Chem. Phys.}  {\bf 112} 7518--7523 (2000).

\bibitem{yamamuro} Yamamuro, O., Tsukushi, I., Lindqvist, A.,
Takahara, S., Ishikawa, M. and Matsuo, T. Calorimetric Study of Glassy
and Liquid Toluene and Ethylbenzene: Thermodynamic Approach to Spatial
Heterogeneity in Glass-Forming Molecular Liquids. {\it
J. Phys. Chem. B} {\bf 102} 1605--1609 (1998).

\bibitem{ngai}  Ngai, K. L. and Yamamuro, O. Thermodynamic fragility 
and kinetic fragility in supercooled liquids: A missing link in molecular
liquids. {\it J. Chem. Phys.} {\bf 111} 10403--10406 (1999).

\bibitem{fsaging} Sciortino, F. and Tartaglia, P. ``Extension of the
Fluctuation-Dissipation theorem to the physical aging of a model
glass-forming liquid'' {\it Phys. Rev. Lett.} (in press). 

\bibitem{starrwater} Starr, F. W., Sastry, S., La Nave, E., Scala, A.,
Sciortino, F. and Stanley, H. E.  ``Thermodynamic and structural
aspects of the potential energy surface of simulated water''
http://arXiv.org/abs/cond-mat/0007487.

\end{thebibliography}
\end{document}